# Linear and Non-linear Estimation Techniques: Theory and Comparison

Raja Manish
*Graduate Student, Aeronautics and Astronautics Engineering*
*Purdue University, West Lafayette, USA*

Under
Inseok Hwang, *Ph.D.*
*Associate Professor, School of Aeronautics and Astronautics*
*Purdue University, West Lafayette, USA*

**The wide application of estimation techniques in system analysis enable us to best determine and understand the history of system states. This paper attempts to delineate the theory behind linear and non-linear estimation with a suitable example for the comparison of some of the techniques of non-linear estimation.**

## Nomenclature

| | | |
|---|---|---|
| $F_x$ | = | probability distribution function |
| $f_x$ | = | probability density function |
| $\omega$ | = | elementary outcomes |
| $\Omega$ | = | collection of points in n-dimensional space |
| x, y, X, Y | = | random variables |
| P | = | probability (0-1) |
| E | = | expectation of variable |
| $m$ | = | mean |
| $P$ | = | covariance matrix |

## I. Introduction

Most of the physical processes that we see around have been developed and represented in a form of some mathematical system model. Use of system models enables us to effectively analyze the past as well as predict the future behavior of the process to some extent. These models can be categorized as deterministic and stochastic models (the term 'stochastic' means random and which involves probability). Deterministic models are easy to represent and compute. Nevertheless, deterministic system do not provide enough information and need for stochastic models becomes important. This can be explained as following:

i. No mathematical system is perfect. Only dominant modes of the system are depicted in the model.
ii. Many effects which has uncertainty are approximated by mathematical model. This loses the accuracy of the result.
iii. Dynamic systems are driven not only by control input but also by disturbances which can neither be controlled nor modelled deterministically.
iv. Sensors do not provide perfect and complete data about a system.

## II. Background of Estimation

### A. Probability theory and Random Variables
*Random Variable* – A real value point function which assigns a real scalar value to each point ω in Ω denoted as X(ω) = x
  Where $\omega$ = elementary outcomes, e.g. $\omega = \{1\}, \{2\}, \{3\}$ etc.

$\Omega$ = Collection of points in n-dimensional space Rn, e.g. $\Omega = \{1, 2, 3 \; etc.\}$
Mathematically it is given as $A = \{\omega: X(\omega) \leq \xi\}$ where $\xi$ is any value on real line

*Probability distribution function* (PDF) – It assigns a probability to each measurable subset of the possible outcomes of a random experiment. In simple words, PDF relates probability P (A) and realizations of random variables. It is defined by

$$F_x(\xi) = P(\{\omega: X(\omega) \leq \xi\})$$

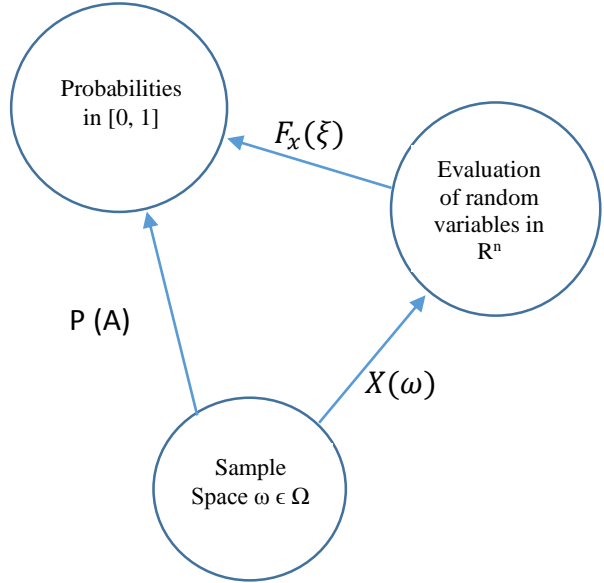

**Fig.1: Probability determination of random variable X**

*Probability density Function* (pdf) – It describes the relative likelihood of a continuous random variable to take on a given value. Mathematically,
If scalar valued function $f_x(.)$ exists such that

$$F_x(\xi_1, \xi_2 \ldots \xi_n) = \int_{-\infty}^{\xi_1} \int_{-\infty}^{\xi_2} \ldots \int_{-\infty}^{\xi_n} f_x(\rho_1, \rho_2 \ldots \rho_n) d\rho_1 d\rho_2 \ldots d\rho_n$$

Or simply $F_x(\xi) = \int_{-\infty}^{\xi} f_x(\rho) \, d\rho$ holds for all $\xi = \begin{bmatrix} \xi_1 \\ \xi_2 \\ . \\ \xi_n \end{bmatrix}$

Then function $f_x(.)$ is called Probability density function of X.

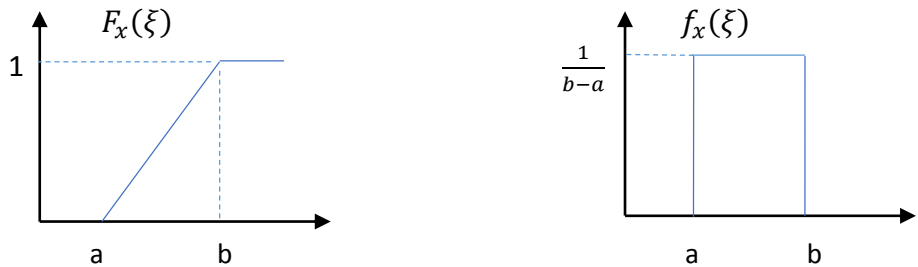

**Fig.2: PDF with its corresponding pdf**

Note: -
i) Pdf exists only when $F_x$ is absolutely continuous. Here $F_x$ is continuous if number of points where it is not differentiable is countable. If such pdf exist then x is called Continuous random variable.

ii) $f_x(\xi) \geq 0$ for all $\xi$

iii) With $F_x$ continuous $P(\{\omega: X(\omega) = \xi_0\}) = F_x(\xi_0) - F_x(\xi_0^-) = 0$
So $P(\{\omega: X(\omega) \in (\xi_1, \xi_2] or [\xi_1, \xi_2]\}) = F_x(\xi_2) - F_x(\xi_1) = \int_{\xi_1}^{\xi_2} f_x(\rho) d\rho$
For infinitesimal interval $\xi_1$ to $(\xi_1 + d\xi_1)$, the integration term, i.e. probability becomes $f_x(\xi)d\xi$.

iv) In case of two random variables X and Y, they are described through joint density function $f_{x,y}(\xi, \rho)$ and marginal density of each variable is given as

$$f_{x|y}(\xi \mid \rho) = \frac{f_{x,y}(\xi, \rho)}{f_y(\rho)}$$

v) Independence of X and Y –
X and Y are independent if

$$P(\{\omega: X(\omega) \in A \text{ and } Y(\omega) \in B\}) = P(\{\omega: X(\omega) \in A\}). P(\{\omega: Y(\omega) \in B\}),$$
$$\text{Or, } F_{x,y}(\xi, \rho) = F_x(\xi). F_y(\rho)$$

Also if the distribution function has well defined derivative then

$$f_{x,y}(\xi, \rho) = f_x(\xi). f_y(\rho)$$

gives the condition for independence (Although independence doesn't require this condition). It has to be noted that there is a difference between Independent and mutually exclusive X and Y.

| Mutually Exclusive | Independent |
|---|---|
| P(A) = 1 implies P(B) = 0 | P(A) gives no information about P(B) |
| P(B) = 1 implies P(A) = 0 | P(A and B) = P(A).P(B) |

Table 1: Mutually exclusive versus Independent

**B. Expectation and Moment of random variables**
*Expected Value* – It is the average value one would obtain over various outcomes of an experiment.
Expected value generates moments of a random variable which are parameters that characterize the distribution or density function. In case of Gaussian random variable, the first two moments can completely describe the distribution or density function.
Let Y be m dimensional vector function of X

$$Y(.) = \theta(X(.)) \text{ Where } \theta \text{ is continuous}$$

Then expectation of Y is given as

$$E[Y] = \int_{-\infty}^{\infty} \theta(\xi) f_x(\xi) d\xi$$

Also, $E[Y] = \int_\Omega \theta(X(\omega)) dP(\omega)$ when density function $f_x(\xi)$ does not exist.

**C. Moments of X**
i) First moment is called as '*Mean*'.

$$m = E[X] = \int_{-\infty}^{\infty} \xi f_x(\xi) d\xi$$

Here m is not random.
ii) Second moment of X is called as '*auto-correction*' matrix when

$\Psi ij = E[XiXj]$ where i-j components of X corresponds to elements of the matrix

$$XX^T = \begin{bmatrix} X_1^2 & X_1X_2 & X_1X_n \\ \vdots & \ddots & \vdots \\ X_nX_1 & \ldots & x_n^2 \end{bmatrix}$$

iii) A second central moment of X is called as '*Covariance matrix*' of X given as

$$P \triangleq E[(X-m)(X-m)^T]$$

Where $m$ = mean
$P$ is symmetric positive semi-definite matrix

iv) Another term known as '*Variance*' is defined as the diagonal components of P, i.e.

$$P_{ii} \triangleq E[(X_i - m_i)^2]$$

Square root of Pii is the Standard Deviation where $P_{ii} \triangleq \sigma_i^2$.

v) Correlation coefficient of Xi and Xj is given as

$$r_{ij} \triangleq \frac{P_{ij}}{\sigma_i \sigma_j}$$

Note:
i) The scalar random variable notations can be similarly applied over n-dimensional random vector.
ii) If vector X and Y are independent, this implies they are uncorrelated.

i.e. $E[XY^T] = E[X].E[Y^T]$.

**D. Gaussian Random Vectors**
It provides an adequate model of the random behavior by many phenomena observed in nature. Also, it yields understandable mathematical models to the base estimators and controllers. It is described through pdf of the form

$$f_x(\xi) = \frac{1}{(2\pi)^{n/2}|P|^{1/2}} e^{\left\{-\frac{1}{2}[\xi-m]^T P^{-1}[\xi-m]\right\}} \qquad \text{\{For Gaussian random vector\}}$$

$$f_x(\xi) = \frac{1}{\sqrt{2\pi P}} e^{\left\{-\frac{1}{2P}[\xi-m]^2\right\}} \qquad \text{\{For scalar Gaussian random variable\}}$$

Where P is positive definite (n x n) matrix which determines size and angular orientation of ellipses of constant likelihood.

**E. Central limit theorem**
If random phenomenon we observe is generated as the sum of effects of many independent random phenomena then the distribution of observed phenomenon approaches a Gaussian distribution. This theorem has a very important consequence as it makes it easy to compute when random phenomena are considered as Gaussian.
Note:
i) All odd central moments of a Gaussian random vector are zero due to symmetry while all even central moments can be expressed in terms of the covariance.
ii) Jointly Gaussian random vectors which are uncorrelated are also independent.

**F. Linear Operation on Gaussian Random Variables**
Linear transformations of Gaussian random variables are also Gaussian random variables. Also, linear combinations of random and non-random Gaussian vectors are also Gaussian random vectors.

If 'X' is Gaussian random n-vector with mean $m_x$ and covariance $P_{xx}$, A is a known matrix then random vector Y defined by $Y = AX$ is Gaussian with mean $Am_x$ and covariance $AP_{xx}A^T$.

If $z = Ax + By + c$,

Then, mean = $Am_x + Bm_x + C$ and

Covariance = $AP_{xx}A^T + AP_{xy}B^T + BP_{yx}A^T + BP_{yy}B^T$

## III. Estimation Theory

### A. Estimation with static linear Gaussian system model
Components of an estimation problem are: -
1. Variables to be estimated $X$, $n$-dimensional
2. Measurements and observations available $Z$, $m$-dimensional
3. The mathematical model describing how the measurements are related to the variables of interest
   $Z = HX + v$, where $v =$ uncertain measurement disturbance
4. Mathematical model of the uncertainties present-
   In $Z = HX + v$, $v$ is the random variable to describe noise corruption whose mean = 0 and covariance = R. Generally v is taken as white Gaussian noise.

   *White Gaussian noise* – A process $X(\cdot,\cdot)$ is a white Gaussian process if for any choice of $t_1, t_1, \ldots t_N \in T$, the N random vectors $X(t_1), \ldots X(t_N)$ are independent Gaussian random vectors.

   $$\Rightarrow P_{xx}(t_i, t_j) = 0 \text{ For } i \neq j$$

5. Performance evaluation criterion to judge which estimation algorithm is 'best'.
   This is done by calculating conditional mean and covariance of X variable given Z as measurements.
   Conditional mean,

   $$\hat{X}^+ = E_x[x|Z = z] = \hat{X}^- + [P^- H^T][HP^- H^T + R]^{-1}[z - H\hat{X}^-]$$

   Conditional covariance,

   $$P^+ = P^- - [P^- H^T][HP^- H^T + R]^{-1}[HP^-]$$

   If we choose gain matrix K = $[P^- H^T][HP^- H^T + R]^{-1}$ then

   $$\hat{X}^+ = \hat{X}^- + K[z - H\hat{X}^-]$$
   $$\text{And } P^+ = P^- - K[HP^-]$$

   We then choose $\hat{X}^+$ as an optimal estimate of the variable of interest. Later we will come to know that form of these equations is same as Kalman filter.

### B. Methods of estimate processing
1. Recursive processing - Here we consider '*a priori*' information about X.
   Given $Z_1(\omega) = z_1$ as measurement and its variance $\sigma_{z_1}^2$, therefore $\hat{X}^+ = z_1$ and $P^- = \sigma_{z_1}^2$.
   Next we consider $Z_2(\omega) = z_2$ and variance $\sigma_{z_2}^2$ as available measurement to be incorporated in the estimate. This way, the process goes on and estimate is updated recursively.

2. Batch processing – Here we do not have '*a priori*' information about X. Hence this is modelled through Gaussian random variable with infinite variance $P^- = \infty$ or $[P^-]^{-1} = 0$.
   Both Recursive and Batch processing yield equivalent results. This equivalence is utilized in design of online estimator. Among these two types, recursive form might be substantially better implementation since partially updated estimated would be preferable to one not updated at all (in case of batch processing).

## IV. Optimal Linear Estimator: Kalman Filter
It is an optimal recursive data processing algorithm. 'Optimal' means it minimizes error in some respect. It processes all available measurements regardless of their precision, to estimate the current value of the variables of interest. Also, it does not require all previous data to be stored and reprocessed every time new measurement is taken. This filter incorporates discrete time measurement samples rather than continuous time inputs. Filter is used as a mean of inferring variables of interest to describe the state of the system from the available data.

Kalman filter combines all available measurement data plus prior knowledge about the system and measuring devices to produce an estimate of the desired variables in such a manner that error is minimized statistically. It propagates the conditional probability density of desired quantities conditioned on knowledge of actual data coming from the measuring devices. This probability density propagation follows three assumptions in order to vastly simplify the mathematics involved:
i. Linear system model – easily manipulated with engineering tools

ii. Noise is white – noise value is not correlated in time and has equal power at all frequencies
iii. Noise is Gaussian – probability density of Gaussian noise takes on the shape of a normal bell shaped curve.

In absence of any higher order statistics than first and second, Gaussian density is the best form to assume. Hence Kalman filter which propagates the first and second order statistics include all information contained in conditional probability density.

**Illustration –**

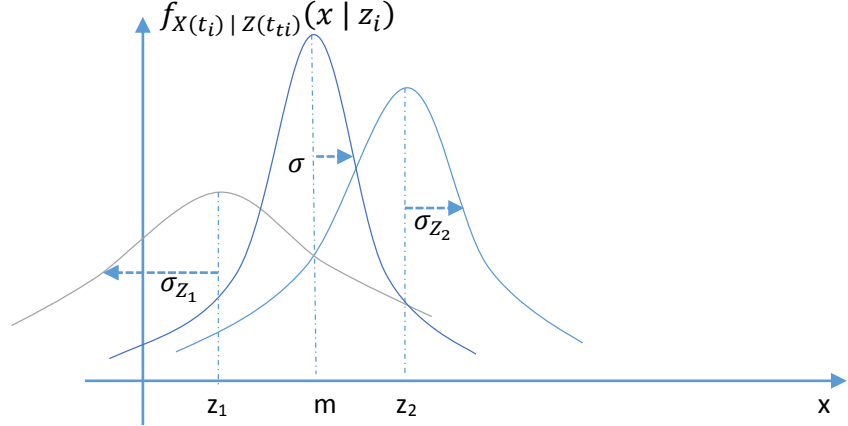

**Fig. 3: Probability density curve showing mean and variance**

Above probability density plot tells us about the probability of being in any location.

1. At time $t_1$ we observe value of the measurement being $z_1$. Also $\sigma_{z_1}$ is the direct measure of uncertainty which is Gaussian. So, larger the $\sigma_{z_1}$, broader will be the probability curve. At this time our best estimate is $\hat{x}(t_1) = z_1$ and variance of error in the estimate is $\sigma_x^2(t_1) = \sigma_{z_1}^2$.

2. At time $t_2 \cong t_1$ we obtain a measurement $z_2$ with a variance $\sigma_{z_2}^2$ where $\sigma_{z_2}^2 < \sigma_{z_1}^2$. Narrower peak means less variance and hence more certainty of the position. To estimate the position we combine the data of both time $t_1$ and $t_2$. We get
Mean
$$m = \frac{\sigma_{z_2}^2 z_1}{\sigma_{z_1}^2 + \sigma_{z_2}^2} + \frac{\sigma_{z_1}^2 z_2}{\sigma_{z_1}^2 + \sigma_{z_2}^2}$$
And variance
$$\frac{1}{\sigma^2} = \frac{1}{\sigma_{z_1}^2} + \frac{1}{\sigma_{z_2}^2}$$

We can see that the new $\sigma^2$ is less than either of $\sigma_{z_1}^2$ and $\sigma_{z_2}^2$, i.e. uncertainty has been decreased.
Therefore, best estimate at time $t_2$,
$$\hat{x}(t_2) = m$$
This is the *maximum likelihood estimate*.
In the Kalman filter form we can write as
$$\hat{x}(t_2) = \hat{x}(t_1) + K(t_2)[z_2 - \hat{x}(t_1)] \qquad \text{Where } K(t_2) = \frac{\sigma_{z_1}^2}{\sigma_{z_1}^2 + \sigma_{z_2}^2}$$
$$\sigma_x^2(t_2) = \sigma_x^2(t_1) - K(t_2)\sigma_x^2(t_1)$$

Therefore, optimal estimate at time $t_2$, $\hat{x}(t_2)$ is equal to best prediction of its value before $z_2$ is taken, $\hat{x}(t_1)$ plus correction term of optimum weighing value times the difference between $z_2$ and best prediction pf its value before it is actually taken.

3. Similar estimate is obtained for next measurement.

The previous illustration is for a static estimation problem. In case of dynamic system we have following relations:

Suppose a motion is of form $\frac{dx}{dt} = u + w$, ($u$ is nominal velocity and $w$ is noise term) and we are at time $t_3^-$, i.e. just before the measurement is taken at time $t_3$. Then Gaussian mean at that time

$$\hat{x}(t_3^-) = \hat{x}(t_2) + u(t_3 - t_2)$$

And corresponding variance

$$\sigma_x^2(t_3^-) = \sigma_x^2(t_2) + \sigma_\omega^2(t_3 - t_2)$$

Now, if we get a measurement value of $z_2$ and variance $\sigma_{z_3}^2$ then we have

$$\text{Mean, } \hat{x}(t_3) = \hat{x}(t_3^-) + K(t_3)[z_3 - \hat{x}(t_3^-)]$$

And variance, $\sigma_x^2(t_3) = \sigma_x^2(t_3^-) - K(t_3)\sigma_x^2(t_3^-)$ where gain $K(t_3) = \frac{\sigma_{x(t_3^-)}^2}{\sigma_x^2(t_3^-) + \sigma_{z_3}^2}$

Depending on the measurement noise variance we may have 3 cases: -
1. Measurement noise variance $\sigma_{z_3}^2$ is large, and then $K(t_3)$ will be small, i.e. an infinitely noisy measurement is totally ignored.
2. Dynamic system noise variance $\sigma_\omega^2$ is large, then $\sigma_x^2(t_3^-)$ will be large and hence $K(t_3)$ will be large, i.e. the measurement is most accurate estimate.
3. Dynamic system noise variance $\sigma_\omega^2$ is small, then $\sigma_x^2(t_3^-) \to 0$ and then $K(t_3) \to 0$. So $\hat{x}(t_3) = \hat{x}(t_3^-)$, i.e. we become confident of our estimate before $z_3$ is measured so measurement can be disregarded.

Note: For non-linear differential equation, we obtain an approximation called the Linearized perturbation equation and further proceed with discretized system.

**Kalman Filter: Algorithm Development Procedure**
1. Define all state definition fields for difference equation $X(k+1) = AX(k) + BU(k) + W$ as well as observation vector $Z = HX + V$:

    A – State transition matrix
    U – Input Vector
    B – Input matrix
    W – Gaussian process noise with mean = 0 and covariance = Q
    H – Observation matrix
    V – Gaussian measurement noise with mean = 0 and covariance = R

2. Define initial state estimate: Xo and covariance Po.
3. Obtain observation and control vectors: Z, U
4. Call the filter to obtain updated state estimate: X, P
5. Go to step 3 and do the iteration

Governing equations:
i) $X(k+1)^- = AX(k) + BU(k)$
ii) $P(k+1)^- = AP(k)A^T + Q$
iii) $K = P^-H^T[HP(k+1)^-H^T + R]^{-1}$
iv) $X(k+1)^+ = X(k+1)^- + K(k+1)[Z(k+1) - HX(k+1)^-]$
v) $P(k+1)^+ = P(k+1)^- - K(k+1)HP(k+1)^-$

## V. Non-Linear Estimation

**A. Extended Kalman Filter (EKF)**
A non-linear stochastic differential equation can be written as

$$\dot{x}(t) = f(x(t), t) + w(t) \text{ Where } f = \text{non-linear function of state}$$
$$w(t) = \text{Zero mean Gaussian noise with spectral density matrix } Q(t)$$

Sampled non-linear measurements

$$Z_k = h_k(X(t_k)) + V_k \text{ For k = 1, 2…}$$

Where, $h_k$ depends on index 'k' and state,
$V_k$ = white random sequence of zero mean Gaussian random variables with covariance matrix $R_k$.

In EKF, state is propagated from $t_{k-1}$ to $t_k$ by integrating from $t_{k-1}$ to $t_k$ and then taking expectation both sides conditioned on all measurements taken up until $t_{k-1}$, then interchanging expectation, integration and differentiating. For general non-linear case we have following

$$\dot{\hat{x}}(t) = f(\hat{x}(t)) \text{ Where } t_{k-1} < t < t_k$$

$$P(t) = F(\hat{x}(t), t)P(t) + P(t)F^T(\hat{x}(t), t) + Q(t) \text{ Where } F(\hat{x}(t), t) \triangleq \frac{\partial f_i(x(t), t)}{\partial x_j(t)}\bigg|_{x(t) = \hat{x}(t)}$$

Above equations are expressions for propagating the conditional mean of state and its associated covariance matrix. They are referred as Extended Kalman Filter (EKF).

Just like the Kalman filter but differently, the EKF measurement update equations are given as

$$\hat{x}_k(+) = \hat{x}_k(-) + K_k[z_k - h_k(\hat{x}_k(-))]$$

$$K_k = P_k^-(-)H_k^T(\hat{x}_k(-))[H_k(\hat{x}_k(-))P_k(-)H_k^T(\hat{x}_k(-)) + R_k]^{-1}$$

$$P_k(+) = [I - K_k H_k(\hat{x}_k(-))]P_k(-)$$

$$\text{Where } H_k(\hat{x}_k(-)) = \frac{\partial h_k(x)}{\partial x}\bigg|_{x = \hat{x}_k(-)}$$

This is the EKF for non-linear systems with discrete measurements. Here $K_k$ is a random variable and $f$ and $h_k$ are linearized about current estimate of $x(t)$ (using $F(\hat{x}(t), t)$ and $H_k(\hat{x}_k(-))$).

Notes:
1. $K_k$ must be computed in real time. $P_k$ is also random depending upon the time history of $\hat{x}_k(t)$.
2. As we can see, state vector is propagated by applying the non-linear relations while the measurement update is done by discrete linear equations. This bring a flaw in the filtering process which we will learn subsequently.

**B. Flaws of EKF**
1. EKF provides only first order approximations to the optimal terms.
2. These approximations can introduce large errors in the true posterior mean and covariance of the transformed Gaussian random vector (GRV).
3. Further this may lead to suboptimal performance or sometimes divergence of the filter.

For these reasons we introduce a different variant of estimator known as Unscented Kalman Filter (UKF).

**C. Unscented Kalman Filter (UKF)**
The state distribution is now specified using minimal set of carefully chosen sample points. These sample points completely capture true mean and covariance of the GRV.

**D. Unscented Transformation**
It is a method for calculating the statistics of a random variable which undergoes a non-linear transformation.

Steps:
1. Let us consider propagating a random variable $x$ with dimension $L$ through a non-linear function $y = g(x)$.
2. Assume x has mean $\bar{x}$ and covariance $P_x$. To calculate statistics of $y$ we form a matrix '$\chi$'(called as '*chi*') of $2L + 1$ sigma vectors $\chi_i$ (with corresponding weight $W_i$) according to the following:

$$\chi_0 = \bar{x}$$

Sigma points:

$$\chi_i = \bar{x} + \left(\sqrt{(L + \lambda)P_x}\right)_i \text{ Where } i = 1, 2\text{… up to L}$$

Also, $\chi_i = \bar{x} - \left(\sqrt{(L+\lambda)P_x}\right)_{i-L}$ for $i = L+1, L+2\ldots$ up to $2L$

Weights:

$$W_0^{(mean)} = \frac{\lambda}{L+\lambda}$$

$$W_0^{(Covariance)} = \frac{\lambda}{L+\lambda} + (1 - \alpha^2 + \beta)$$

$$W_i^{(mean)} = W_i^{(Covariance)} = \frac{1}{2(L+\lambda)} \text{ For } i = 1,2\ldots \text{ up to } 2L$$

Where  $\lambda = \alpha^2(L+k) - L$
$L$ = Scaling parameter
$\alpha$ = Spread of sigma points around $\bar{x}$ (a small value e.g. $10^{-3}$)
$k$ = Secondary scaling parameter (usually 0)
$\beta$ = To incorporate prior knowledge of distribution (for Gaussian, $\beta = 2$ is optimal)

3. These sigma vectors are propagated through non-linear function $Y_i = g(\chi_i), i = 0, 1 \ldots 2L$.
4. Mean and covariance for $y$ are approximated using weighted sample mean and covariance of the posterior sigma points.
Mean:

$$\bar{y} \approx \sum_{i=0}^{2L} W_i^{(m)} Y_i$$

$$P_y \approx \sum_{i=0}^{2L} W_i^{(c)} \{Y_i - \bar{y}\}\{Y_i - \bar{y}\}^T$$

5. Finally we have following state estimation algorithm:
   a) Redefine UKF state random vector as concatenation of the original state and noise variables.

   $$X_k^a = [X_k^T \quad v_k^T \quad n_k^T]^T \text{ Where } v = \text{process noise}$$

   $$n = \text{Measurement noise}$$

   b) Calculate sigma points-

   $$\chi_{k-1}^a = \left[\hat{x}_{k-1}^a \quad \hat{x}_{k-1}^a - \left(\sqrt{(L+\lambda)P_{k-1}^a}\right)_{i=0 \text{ to } L} \quad \hat{x}_{k-1}^a + \left(\sqrt{(L+\lambda)P_{k-1}^a}\right)_{i=(L+1) \text{ to } 2L}\right]$$

   c) Time update (propagation)

   $$\chi_{k|k-1}^x = F[\chi_{k-1}^x \quad \chi_{k-1}^v] \qquad \{\text{Based on the non-linear equations}\}$$

   $$\hat{x}_k^- = \sum_{i=0}^{2L} W_i^{(m)} \chi_{i,k|k-1}^x$$

   $$P_k^- = \sum_{i=0}^{2L} W_i^{(c)} \{\chi_{i,k|k-1}^x - \hat{x}_k^-\}\{\chi_{i,k|k-1}^x - \hat{x}_k^-\}^T$$

   $$Y_{k|k-1} = H[\chi_{k|k-1}^x \quad \chi_{k-1}^n] \qquad \{\text{Based on the measurement equations}\}$$

   $$\hat{y}_k^- = \sum_{i=0}^{2L} W_i^{(m)} Y_{i,k|k-1}$$

   d) Measurement update equations

   $$P_{\tilde{y}_k \tilde{y}_k} = \sum_{i=0}^{2L} W_i^{(c)} \{Y_{i,k|k-1} - \hat{y}_k^-\}\{Y_{i,k|k-1} - \hat{y}_k^-\}^T \text{ Where } \tilde{x}_k = x_k - \hat{x} = \text{error in estimate}$$

   $$P_{\tilde{x}_k \tilde{y}_k} = \sum_{i=0}^{2L} W_i^{(c)} \{\chi_{i,k|k-1} - \hat{x}_k^-\}\{Y_{i,k|k-1} - \hat{y}_k^-\}^T \qquad \text{Where } \tilde{y}_k = y_k - \hat{y}$$

$$K = P_{\tilde{x}_k \tilde{y}_k} P_{\tilde{y}_k \tilde{y}_k}^{-1}$$

$$\hat{x}_k = \hat{x}_k^- + K(y_k - \hat{y}_k^-)$$

$$P_k = P_k^- - K P_{\tilde{y}_k \tilde{y}_k} K^T$$

| **Extended Kalman Filter (EKF)** | **Unscented Kalman Filter (UKF)** |
|---|---|
| The state distribution is approximated by a Gaussian random variables (GRV) | The state distribution though approximated by GRV, but now represented using minimal set of carefully chosen sample points which completely captures mean and covariance. |
| State is then propagated analytically through first order linearization of the non-linear system. This can introduce large error in true posterior mean and covariance of the transformed GRV. | State is propagated through the true non-linear system, which captures posterior mean and covariance accurately to the 3rd order (Taylor series expansion). |
| It may lead to suboptimal performance and sometime divergence of the filter. | UKF is better in performance with same order of computation complexity as that of the EKF. |
| Calculation of Jacobian or Hessian required over non-linear system | No such calculations are required. |

Table 2: Comparison between Non-linear Estimator, EKF and UKF

## VI. Example Problem

In this example (See 6-1-2[2]) we consider a problem of tracking a body falling freely through the atmosphere. The motion is modeled in one dimension by assuming the body falls in straight line, directly towards the tracking radar.

State variables are $x_1 = x$; $x_2 = \dot{x}$; $x_3 = \beta$, position, velocity and ballistic co-efficient respectively.

The equations of motion are given by

$$\dot{x}_1 = x_2$$

$$\dot{x}_2 = \frac{\rho x_2^2}{2 x_3} - g \text{ Where } \rho = \rho_0 e^{-\frac{x_1}{k_\rho}}$$

$$\dot{x}_3 = 0$$

Where $\rho$ = exponential approximation of air density with respect to altitude

$\rho_0$ = Mass density of air at mean sea level = $2.377 \times 10^{-3} \frac{slugs}{ft^3}$ or $lb.s^2.ft^{-4}$

$k_\rho$ = A constant, known as scaled height which is approximately 22000 ft.

The initial values of states are $\begin{bmatrix} 10^5 \: ft \\ -6000 \: ft/s \\ 2000 \end{bmatrix}$ and variances $\begin{bmatrix} 500 \: ft^2 \\ 2 \times 10^4 \: (ft/s)^2 \\ 2.5 \times 10^5 \end{bmatrix}$

The non-linear continuous time system is discretized with time interval of one second using "Euler Maruyama method"[4].

**Result**
Following is the result obtained for the errors in position estimates using Linearized Kalman filter, the EKF and the UKF:

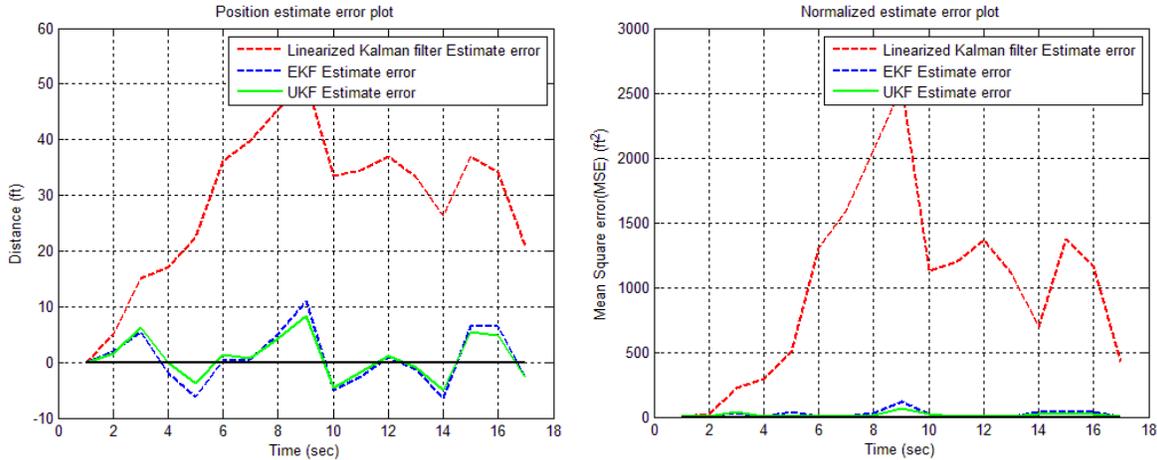

**Fig. 4: (a) Linearized KF, EKF and UKF estimate error characteristics**
**(b) Normalized estimate error plot**

The Matlab simulation (fig. 4) results in different plot each time the code is executed. This is due to the random function coded with the measurement algorithm. In the above plot I have assumed that the system model for Linearized KF (LKF) remains constant throughout the algorithm since only the initial values of states are used to compute the system model (and hence system model is not updated in each step) whereas in EKF the linearized system model is numerically computed using latest estimate of the state at each time step. In contrast, UKF follows its algorithm based on non-linear dynamics without linearizing the system. Another thing which can be noticed is that the LKF plot seems to converge in the later time steps. This is probably because the algorithm gives more weight on increasingly accurate measurements of the position when compared to the estimate as the object approaches the tracking radar.

It is evident that the UKF shows a better performance than both LKF and EKF owing to its computation over 3rd order system. It was able to retain the accuracy to a high extent. Also, unlike LKF and EKF there is no need for linearization. In contrast, the LKF shows a large amount of error in the estimate making it almost unsuitable for the intended non-linear system.

## VII. Conclusion

Estimators plays a very important role in understanding the history of the system. The Kalman filter is indeed the best estimator for the linear system. Considering the non-linear nature of almost all the processes due to disturbances and noise of the sensors, the Kalman filter is generally not employed in generic form. Instead, the linearized KF, EKF and UKF are adopted for this purpose. As we have seen, the LKF, EKF and UKF shows different characteristic results. The type of estimator to use depends on the purpose, complexity and accuracy required for a given system. Also, systems where an error is the desired quantity, these estimators may be modified to obtain the result but effectively all of the estimators works on the similar principle.

Further, there are more variants of the filters like the Particle filter in which we try to estimate the posterior density of the samples which are set of particles. In future, I intend to include Particle filters in the study including its various types such as Gaussian Particle filter, unscented particle filter and other recursive Bayesian estimators.

# Appendix

**Matlab Code**

```matlab
%~~~~~~ Application of Linear and Non-linear Estimation~~~~~~

%Linearized Kalman filter, Extended Kalman Filter and
%Unscented Kalman filters

%~~~~~~~By R Manish~~~~~~%

clear all;clc;
syms rho g k_rho r beta x0 xdot0 p11 p22 p33 x1 x2 x3

%define some constants
rho = 2.377*10^-3;   %slugs/ft^3
g = 32.4;            %ft/sec^2
k_rho = 22000;       %ft

r = 100;    %measurement noise cov
Xo = [10^5; -6000; 2000];    %Initial value of states
p11 = 500;
p22 = 2*10^4;
p33 = 2.5*10^5;
Po = [p11 0 0;0 p22 0;0 0 p33];       %Initial Covariance matrix

stdev.process = [2;5;8]; %Standard deviation of the process
%Process noise covariance, written explicitly for ease of understanding
Q = [stdev.process(1)^2 stdev.process(1)*stdev.process(2) stdev.process(1)*stdev.process(3);
     stdev.process(2)*stdev.process(1) stdev.process(2)^2 stdev.process(2)*stdev.process(3);
     stdev.process(3)*stdev.process(1) stdev.process(3)*stdev.process(2) stdev.process(3)^2];

x = [x1;x2;x3];      %State variables
h = @(x) x(1);       %Observation function
f = [x(1)+x(2); x(2)+((rho*exp(-x(1)/k_rho)*x(2)^2)/(2*x(3)))-g; 2000+0];    %Non-linear system definition
F = @(x) [x(1)+x(2); x(2)+((rho*exp(-x(1)/k_rho)*x(2)^2)/(2*x(3)))-g; 2000+0];  %Same system declared for function calls
A = jacobian(f,x);       %Find Jacobian of nonlinear function

disp('Time steps:')
t = 18
B = [0;0;0];
u = 0;
```

```matlab
%Generate noises
numsamples = t;
sample.pos = stdev.process(1).*randn(numsamples,1);
sample.vel = stdev.process(2).*randn(numsamples,1);
sample.accl = stdev.process(3).*randn(numsamples,1);
noise.process = transpose([sample.pos, sample.vel, sample.accl]);
noise.mes = sqrt(r).*randn(numsamples,1);

%create measurements
x_tru = [];
z = [];
S = Xo;

for i = 1:t
    x_tru(:,end+1) = S;
    x1=x_tru(1,end);x2=x_tru(2,end);x3=x_tru(3,end);
    z(:,end+1) = h(x_tru(:,end)) + noise.mes(i);
    Fx = eval(f);
    S = Fx + B*u + noise.process(:,i);
end

%% Linearized kalman Filter algorithm
disp('Executing Linearized Kalman Filter')
x_k = Xo;
P_k = Po;
K_k=[];
x_k_hat=[];

x1 = Xo(1);x2 = Xo(2);x3 = Xo(3);
A_lin = eval(A);
H_k = [1 0 0];

for i = 1:t-1
    %Propagation
x_k = A_lin*x_k + B*u;
P_k = A_lin*P_k*A_lin' + Q;
K_k = P_k*H_k'*inv(H_k*P_k*H_k'+r);
    %Update after measurement
x_k = x_k + K_k*(z(:,i+1) - H_k*x_k);
P_k = P_k - K_k*H_k*P_k;
x_k_hat(:,end+1) = x_k(:,end);
end
```

```matlab
disp('....done')

%% EKF Algorithm
disp('Executing EKF')
x = Xo;
P = Po;
H = [1 0 0];
Kx=[];
x_hat=[];

for i = 1:t-1
    %Propagate states
    x1=x(1,end);
    x2=x(2,end);
    x3=x(3,end);
    A_est = eval(A);
    x = (F([x(1,end);x(2,end);x(3,end)]));   %Predicted x
    P = A_est*P*A_est' + Q;       %Predicted P
    
    %Update the states
    Kk = P*H'*inv(H*P*H'+r);
    x = x + Kk*(z(:,i+1) - h(x));    %Estimate x
    P = P - Kk*H*P;                  %Estimate P
    x_hat(:,end+1) = x(:,end);
end
disp('....done')

%% UKF Algorithm
disp('Executing UKF')
x_u = Xo;
P_u = Po;
x_u_hat=[];

L=numel(x_u);                               %numer of states
m=1;                                        %numer of measurements
alpha=1e-3;                                 %default, tunable
ki=0;                                       %default, tunable
beta=2;                                     %default, tunable
lambda=alpha^2*(L+ki)-L;                    %scaling factor
c=L+lambda;                                 %scaling factor
Wm=[lambda/c 0.5/c+zeros(1,2*L)];           %weights for means
Wc=Wm;
Wc(1)=Wc(1)+(1-alpha^2+beta);               %weights for covariance
```

```matlab
c=sqrt(c);

Xkmin = [];
Ykmin = [];

for i = 1:t-1
    %Calc Sigma Points
    A_u = c*chol(P_u);                      %Diagonalizing
    Y = x_u(:,ones(1,numel(x_u)));
    X = [x_u Y+A_u Y-A_u];         %Sigma points
    
    %Unscented Transformation(UT) of process
    Lx = size(X,2);            %Size of sigma matrix
    Yz = zeros(3,1);           %Set mean matrix
    Yy = zeros(3,Lx);
    
    for j = 1:Lx
        Yy(:,j) = F(X(:,j));     %Transformed sampling points
        Yz = Yz+Wm(j)*Yy(:,j);   %Weighted mean
    end
    Xkmin(:,end+1) = Yz;         %Store transformed mean as X-
    
    Yerr = Yy - Yz(:,ones(1,Lx));  %Deviation (Xmin-X)
    P1 = Yerr*diag(Wc)*Yerr'+Q;    %Transformed covariance
    
    %UT of Measurement
    Lx = size(Yy,2);
    Zz = zeros(1,1);
    
    for j = 1:Lx
        Zy(:,j) = h(Yy(1,j));    %Transformed Measurement sampling points
        Zz = Zz+Wm(j)*Zy(:,j);   %Weighted measurement mean
    end
    Ykmin(:,end+1) = Zz;         %Store measurement as Y-
    
    Zerr = Zy - Zz(:,ones(1,Lx));    %Deviation (Ymin-Y)
    P2 = Zerr*diag(Wc)*Zerr'+r;      %Measurement covariance
    
    P12 = Yerr*diag(Wc)*Zerr';       %Cross covariance
    K_u = P12*inv(P2);                %Kalman gain
```

```matlab
    x_u = Xkmin(:,end)+K_u*(z(:,i+1)-Ykmin(:,i));    %z(i+1) since Ist measurement is the initial value of position.
    P_u = P1-K_u*P2*K_u';           %Covariance update
    
    x_u_hat(:,end+1) = x_u(:,end);      %Mean update
end
disp('...done')

%% Plot results

figure(1)
plot(1:t-1,x_k_hat(1,:)'-z(1,2:end)','r--',1:t-1,x_hat(1,:)'-z(1,2:end)','b--',1:t-1,x_u_hat(1,:)'-z(1,2:end)','g-',1:t-1,z(1,2:end)'-z(1,2:end)','k','LineWidth',2);
grid on
title('Position estimate error plot');
ylabel('Distance (ft)');
xlabel('Time (sec)');
legend('Linearized Kalman filter Estimate error', 'EKF Estimate error', 'UKF Estimate error');

%%
figure(2)
for i = 1:t-1
    X_k_mse(:,i) = (x_k_hat(1,i)'-z(1,i+1)')^2;
    X_mse(:,i) = (x_hat(1,i)'-z(1,i+1)')^2;
    X_u_mse(:,i) = (x_u_hat(1,i)'-z(1,i+1)')^2;
    z_mse(:,i) = (z(1,i+1)'-z(1,i+1)')^2;
end
plot(1:t-1,X_k_mse,'r--',1:t-1,X_mse,'b--',1:t-1,X_u_mse,'g-',1:t-1,z_mse,'k','LineWidth',2);
grid on
title('Normalized estimate error plot');
ylabel('Mean Square error(MSE) (ft^2)');
xlabel('Time (sec)');
legend('Linearized Kalman filter Estimate error', 'EKF Estimate error', 'UKF Estimate error');
```

## Acknowledgments

I would like to thank Dr. Inseok Hwang for providing me with this opportunity of work with him. I would also like to thank Sangjin Lee at Hybrid Systems Laboratory (HSL), Purdue University, for guiding me throughout the learning process. Their timely help was highly beneficial and it was an immense intellectual experience working with them.